\documentclass[12pt]{article}  

\def\[{\left\lbrack}
\def\]{\right\rbrack}

\def\({\left(}
\def\){\right)}
\newcommand{\be}{\begin{equation}}
\newcommand{\ee}{\end{equation}}
\newcommand{\ea}{\end{eqnarray}}
\newcommand{\ba}{\begin{eqnarray}}

\newcommand{\ep}{{\epsilon}}

\usepackage{graphicx}

\begin{document}

\title{Noncommutative cosmological model in the presence of a phantom fluid}

\author{G. Oliveira-Neto and A. R. Vaz\\
Departamento de F\'{\i}sica, \\
Instituto de Ci\^{e}ncias Exatas, \\ 
Universidade Federal de Juiz de Fora,\\
CEP 36036-330 - Juiz de Fora, MG, Brazil.\\
gilneto@fisica.ufjf.br, afonsoricardovaz@gmail.com}

\maketitle

\begin{abstract}
We study noncommutative classical Friedmann-Robertson-Walker cosmological models. The constant 
curvature of the spatial sections can be positive ($k=1$), negative ($k=-1$) or zero ($k=0$). 
The matter is represented by a perfect fluid with negative pressure, phantom fluid, which 
satisfies the equation of state $p =\alpha \rho$ , with $\alpha < - 1$, where $p$ is the pressure 
and $\rho$ is the energy density. We use Schutz's formalism in order to write the perfect 
fluid Hamiltonian. The noncommutativity is introduced by nontrivial Poisson brackets
between few variables of the models. In order to recover a description in terms of commutative 
variables, we introduce variables transformations that depend on a noncommutative parameter ($\gamma$). 
The main motivation for the introduction of the noncommutativity is trying to explain the present 
accelerated expansion of the universe. We obtain the dynamical equations for these models and solve them.
The solutions have four constants: $\gamma$, a parameter associated with the fluid energy $C$, $k$, 
$\alpha$ and the initial conditions of the models variables. For each value of $\alpha$, we obtain 
different equations of motion. Then, we compare the evolution of the universe between the present 
noncommutative models and the corresponding commutative ones ($\gamma \to 0$). The results show that 
$\gamma$ is very useful for describing an accelerating universe. We estimate the value of $\gamma$, 
for the present conditions of the Universe. Then, using that value of $\gamma$, in one of the
noncommutative cosmological models, we compute the amount of time this universe would take to reach the {\it big rip}.
\end{abstract}


\section{Introduction}
\label{sec:intro}

One of the major scientific discoveries of the last century was made in 1998, when
two teams of astronomers observing distant supernovas concluded that our
Universe is expanding in an accelerated rate \cite{expansion}. That amazing discovery
took the scientific community by surprise and since then many different explanations
have appeared in the literature. Many physicists believe, today, that the best explanation
for the accelerated expansion of our Universe consists in the presence of a previously
unknown type of energy. That energy, called: {\it dark energy}\cite{turner}, has properties very different from the usual 
one and should correspond to approximately 74\% of the total matter/energy content of the Universe\cite{caldwell}.
It may be represented by a perfect fluid with equation of state: $p/\rho = \alpha < -1/3$, 
where $\alpha$ is a constant which defines the fluid, $p$ is the fluid pressure and $\rho$ its
density\cite{odintsov}. The first property that makes it very different from usual perfect fluids
comes from the equation of state: its pressure is negative. Depending on the exact value of
$\alpha$ that fluid may violate one or more of the four energy conditions and the resulting
space-time solution may develop one or more of the four types of finite-time future singularities\cite{odintsov}.
We may mention some of the dark energy candidates: cosmological constant, quintessence, quintom, K-essense, phantom fluid,
Chaplygin gas\cite{Mli}. In the present work, we are going to consider as the matter/energy content of our model
a phantom fluid. That fluid has an equation of state with $\alpha < -1$. It violates all four energy conditions
and gives rise to a space-time solution which develops a finite-time future singularity called {\it big rip}\cite{odintsov}.
That singularity appears in a finite time ($t_{br}$), after the beginning of expansion. The scale factor, the fluid
energy density and pressure, all diverge as the time approaches $t_{br}$\cite{mcinnes}. Present day observations, do not 
discard a fluid with an equation of state like the phantom fluid \cite{odintsov}.
As a matter of completeness, we mention that another important explanation for the
present expansion of our Universe considers that general relativity is not the
correct gravity theory. Therefore, it has to be modified. 
For a review on this important field of research see Ref.\cite{odintsov1}.

One of the first ideas introduced, in order to eliminate the divergences in the early days
of quantum field theories, was noncommutativity between spacetime coordinates\cite{snyder}.
The main idea was that, the noncommutativity would induce an uncertainty relation
between the spacetime coordinates. In your turn, that uncertainty relation, would cause the
spacetime points to be replaced by infinitesimal areas of the Planck area order.
Eliminating, in that way, divergent quantities obtained as the result of calculations in
specific spacetime points.
Recently, the interest in those ideas of noncommutativity between spacetime coordinates were 
renewed due to some important results obtained in superstring, membrane and $M$-theories.
For a review on those important results see Ref.\cite{szabo}.
In the past few years, the role played by noncommutativity in different areas of physics has 
been extensively investigated \cite{banerjee}. 
One important arena where noncommutative (NC) ideas may play an important role 
is cosmology. In the early stages of its evolution, the Universe may have had
very different properties than the ones it has today. Among those properties
some physicists believe that the spacetime coordinates were subjected to a 
noncommutative algebra. Inspired by these ideas some researchers have considered 
such NC models in quantum cosmology 
\cite{garcia,nelson,barbosa,gil1}. It is also possible that some residual NC contribution 
may have survived in later stages of our Universe.  
Based on these ideas some 
researchers have proposed some NC models in inflationary cosmology in order to explain 
some intriguing results observed, in the cosmic microwave background radiation (CMB), by different 
sources like: the Planck satellite, the WMAP, BAO and high-l ACT/SPT temperature data. Such as a 
running spectral index of the scalar fluctuations and an anomalously low quadrupole and octopole 
of CMB angular power spectrum \cite{brandenberger}. 
Another relevant application of NC ideas in semi-classical and classical cosmology 
is the attempt to explain the present accelerated expansion of our Universe 
\cite{pedram,obregon,gil,gil2}.

In the present work, we would like to contribute to the investigation on the importance of 
noncommutativity as a possible mechanism to explain the present expansion of the Universe. 
In this way, we study the NC version of a classical
cosmology model. The model has a Friedmann-Robertson-Walker (FRW) geometry, the 
matter content is a phantom fluid and the spatial sections may have 
negative, positive or zero constant curvatures. We work in the Schutz's variational 
formalism \cite{schutz,germano1}. The noncommutativity is
obtained by imposing deformed Poisson brackets between certain canonical variables. 
In fact, the present work is an extension of a previous work\cite{gil2}, where the authors
consider the same NC model coupled to perfect fluids of radiation and dust.
Initially, we derive the scale factor dynamic equations for the general situation, without 
specifying the value of the parameter $\alpha$, which specifies the phantom fluid, or the 
curvature of the spatial sections. Next, we study the scale factor behaviors, for several different
values of $\alpha$ and all possible values of $k$. We compare them with the corresponding 
commutative model. We obtain NC models that may describe the present expansion of our Universe, 
in a better way than the corresponding commutative model. The noncommutativity that we are about to 
propose is not the typical noncommutativity between standard spacetime coordinates. 
In the geometrodynamics formulation of general relativity \cite{wheeler},
the arena in which the classical dynamics takes place is the superspace. 
It is the space of all three-metrics and matter field configurations on a three-surface. In the 
models we are considering here, due to the symmetry of the metric, we have a very simple superspace, 
called minisuperspace, with a small number of `coordinates'. Therefore, in all works done so far in this 
area, the motivation is to extend the usual noncommutativity between standard spacetime coordinates 
to minisuperspace `coordinates' \cite{garcia}. The noncommutativity between minisuperspace
`coordinates' have been studied at the quantum and classical levels. At the quantum
level in Refs.\cite{garcia,nelson,barbosa,gil1} and at the semi-classical and classical levels in
Refs.\cite{pedram,obregon,gil,gil2}. 

In Section \ref{sec:general}, we introduce the NC model for a generic
phantom fluid and derive the coupled system of differential equations for the variables. In
Section \ref{sec:radiation}, we apply the general formalism for several specific cases of phantom
fluids. We solve the system of differential equations and obtain the scale factor as a function
of the time coordinate and few parameters, including the NC parameter $\gamma$. We analyze all
possibles behaviors of the solutions, including a comparison with the solutions to the corresponding 
commutative model, paying special attention for those representing expansion. In Section \ref{sec:endofuniverse},
we give estimates for the NC parameter $\gamma$. Then, using those values of $\gamma$, in one of the
NC cosmological models, we compute the amount of time those universes would take to reach the {\it big rip}.
Finally, in Section \ref{sec:conclusions}, we comment on the most important results of the present paper.

\section{The noncommutative model for a generic phantom fluid}
\label{sec:general}

The FRW cosmological models are characterized by the
scale factor $a(t)$ and have the following line element,
\begin{equation}  
\label{1}
ds^2 = - N^2(t) dt^2 + a^2(t)\left( \frac{dr^2}{1 - kr^2} + r^2 d\Omega^2
\right)\, ,
\end{equation}
where $d\Omega^2$ is the line element of the two-dimensional sphere with
unitary radius, $N(t)$ is the lapse function and $k$ gives the type of
constant curvature of the spatial sections. It may assume the values $k=-1, 1, 0$ 
and we are using the natural
unit system, where $c=G=1$. The matter content of the model is
represented by a perfect fluid with four-velocity $U^\mu = N(t)\delta^{\mu}_0$
in the comoving coordinate system used. The total energy-momentum tensor 
is given by,
\begin{equation}
T_{\delta \nu} = (\rho+p)U_{\delta}U_{\nu} + p g_{\delta \nu}\, ,  
\label{2}
\end{equation}
where $\rho$ and $p$ are the energy density and pressure of the fluid,
respectively. Here, we assume the following state equation of the fluid, $p = \alpha\rho$, where $\alpha < -1$ is a constant
which defines the phantom fluid.

In the present work, we obtained the perfect fluid Hamiltonian using
the Schutz's variational formalism. In this formalism \cite{schutz}, the four-velocity ($U_\nu$) of the fluid is expressed in terms of 
six thermodynamical potentials ($\mu$, $\ep$, $\zeta$, $\beta$, $\theta$, $S$), in the following way,

\be
\label{2.1}
U_\nu = \frac{1}{\mu}\left(\ep_{,\nu}+\zeta\beta_{,\nu}+\theta S_{,\nu}\right).
\ee
Where $\mu$ is the specific enthalpy, $S$ is the specific entropy,
$\zeta$ and $\beta$ are connected with rotation and are absent of FRW models and, 
finally, $\ep$ and $\theta$ have no clear physical meaning. The four-velocity is subject to the normalization condition,

\be
\label{2.2}
U^\nu U_\nu = -1.
\ee

The starting point, in order to write the Hamiltonian of the model, is the action ($\mathcal{S}$) for gravity plus perfect fluid, which in this formalism is written as,

\be
\label{2.3}
\mathcal{S} = \int d^4x\sqrt{-g}(R + 16\pi p),
\ee
where $g$ is the determinant of the metric, $R$ is the curvature scalar and $p$ is the fluid pressure. The last term of (\ref{2.3}) 
represents the matter contribution to the total action. Introducing the metric (\ref{1}) in the action (\ref{2.3}), using the 
geometrodynamics formulation of general relativity \cite{wheeler}, Eqs. (\ref{2.1}) and (\ref{2.2}), the state equation of the fluid, the first law 
of thermodynamics and after some thermodynamical considerations, the action takes the form \cite{germano1},

\be
\label{2.4}
\mathcal{S}=\int dt\left[-6\frac{\dot{a}^2a}{N} + 6kNa + 
N^{-1/\alpha}a^3\frac{\alpha(\dot{\ep}+\theta\dot{S})^{1+1/\alpha}}{(\alpha+1)^{1+1/\alpha}}e^{-S/\alpha}\right].
\ee
From this action, we may obtain the Lagrangian density of the model and write, with the aid of the geometrodynamics formulation
of general relativity, its associated superhamiltonian,

\be
\label{2.5}
N{\mathcal{H}}=N\left(-\frac{P_{a}^2}{24a} - 6ka + P_{\ep}^{\alpha+1}a^{-3\alpha}e^S\right),
\ee
where $P_a = -12\dot{a}a/N$ and $P_\ep = N^{-1/\alpha}a^3(\dot{\ep}+\theta\dot{S})^{(\alpha+1)^{-1/\alpha}/\alpha}e^{-S/\alpha}$.
We may further simplify the superhamiltonian (\ref{2.5}), by performing the following canonical transformations \cite{germano1},

\be
\label{2.6}
T = -P_S e^{-S}P_\ep^{-(\alpha+1)},\quad P_T = P_\ep^{\alpha+1}e^S,\quad \bar{\ep} = \ep-(\alpha+1)\frac{P_S}{P_\ep},\quad \bar{P_\ep} = P_\ep,
\ee
where $P_S = \theta P_\ep$. With these transformations the superhamiltonian (\ref{2.5}) takes the form,

\begin{equation}
N {\mathcal{H}}= -\frac{P_{a}^2}{24} - 6ka^2 + a^{1-3\alpha}P_{T},  
\label{3}
\end{equation}
where $P_{a}$ and $P_{T}$ are the momenta canonically conjugated to $a$ and 
$T$, the latter being the canonical variable associated to the fluid.
Here, we are working in the conformal gauge, where $N = a$.

In order to introduce the noncommutativity in the model, we start considering,
initially, that the total Hamiltonian of the model has the same functional form 
as (\ref{3}). But now it is written in terms of NC variables,
\begin{equation}
N_{nc} {\mathcal{H}}_{nc}= -\frac{P_{anc}^2}{24} - 6ka_{nc}^2 + a_{nc}^{1-3\alpha}P_{Tnc},  
\label{3,5}
\end{equation}
Then, we propose that the noncommutative variables of the model \\
$\{a_{nc}, P_{anc}, T_{nc}, P_{Tnc}\}$ satisfy the following deformed Poisson brackets (PBs):
\ba
\label{4}
\left\{a_{nc},T_{nc}\right\}=\left\{P_{anc},P_{Tnc}\right\}=0,\\
\label{4.1}
\left\{a_{nc},P_{anc}\right\}=\left\{T_{nc},P_{Tnc}\right\}=1,\\
\label{4.2}
\left\{a_{nc},P_{Tnc}\right\}=\left\{T_{nc},P_{anc}\right\}=\gamma,
\ea 
in which $\gamma$ is the NC parameter. It is important to notice that this is
not the only possible deformed PBs one may propose, for the present model. 

In Ref. \cite{gil} the authors considered a very similar classical, noncommutative, FRW model coupled to a perfect fluid, 
in the presence of a cosmological constant. The only differences between our NC model and the NC model in Ref. \cite{gil} are the choices of deformed PBs and the 
presence of a cosmological constant in their model.
In their choice of deformed PBs, they made the two PBs in Eq. (\ref{4}) different from zero, instead of the two PBs in Eq. (\ref{4.2}). 
Therefore, since one of our motivations is investigating possible differences among different 
deformed PBs choices the only possibility, that does not include any of the PBs in Eq. (\ref{4}), was to make the two PBs in Eq. (\ref{4.2}) different from zero. For
simplicity we make them equal to the same NC parameter.
As we mentioned above, the present work is an extension of a previous work\cite{gil2}.
There, the authors consider the same NC model described here coupled to perfect fluids of radiation and dust 
and they also make a detailed comparison between the present NC model and the one introduced in Ref. \cite{gil},
for those two types of fluids. Unfortunately, here, we shall not be able to compare our results with the ones of
Ref. \cite{gil} because, there, the authors did not consider a model with a phantom perfect fluid. We shall
leave it for a future work.

We would like to describe those models in terms of usual commutative variables, which satisfy the usual PBs.
Because it is simpler to deal with that kind of variables.
Following the literature of NC theories it is possible to achieve that by introducing a set of 
coordinate transformations from the NC variables to new commutative ones. Those type of 
transformations were first introduced in Refs. \cite{susskind} and sometimes are called Bopp shift \cite{zachos}.
Due to our choice of deformed PBs (\ref{4.2}), the more general transformations, to first order in $\gamma$, leading
from the NC variables to new commutative ones, are given by,
\ba
\label{5}
a_{nc}\rightarrow a_c + \frac{\gamma}{2}T_c,\nonumber\\
P_{anc}\rightarrow P_{ac} + \frac{\gamma}{2}P_{Tc},\nonumber\\
T_{nc}\rightarrow T_c + \frac{\gamma}{2}a_c,\\
P_{Tnc}\rightarrow P_{Tc} + \frac{\gamma}{2}P_{ac},\nonumber
\ea
where the commutative variables have $c$ labels. It is important to notice that
if we introduce the noncommutative variables Eq. (\ref{5}), in the deformed PBs Eq. (\ref{4}-\ref{4.2}) 
and use the usual PBs among the commutative variables, they are satisfied to first 
order in $\gamma$. Another important motivation to use those commutative variables, is that, the metric for those models
may be written in terms of them as,
\begin{eqnarray}  
\label{5,5}
ds^2 & = & - \left(a_c(t) + \frac{\gamma}{2}T_c(t)\right)^2 dt^2 \nonumber\\
& + & \left(a_c(t) + \frac{\gamma}{2}T_c(t)\right)^2\left( \frac{dr^2}{1 - kr^2} + r^2 d\Omega^2\right)\, .
\end{eqnarray}
For $\gamma = 0$, this metric reduces to Eq. (\ref{1}), in the gauge $N = a$. Observing the metric Eq. (\ref{5,5}), we notice that the dynamics of
world lines separations between two different times is given by the NC scale factor,
\begin{equation}
\label{5,55}
a_{nc}(t) = a_c(t) + \frac{\gamma}{2}T_c(t).
\end{equation}
Therefore, in our study of the dynamics of the models described by the metric Eq. ({\ref{5,5}), we must compute the NC scale factor given by Eq. (\ref{5,55}).
Since, all quantities in that metric Eq. ($\ref{5,5})$ are
commutative, we can treat those models using the usual general relativity methods. In particular, if we write the conservation equation for the
fluid stress-energy tensor Eq. (\ref{2}), for the metric Eq. (\ref{5,5}), we obtain the following relationship between the fluid density and
the NC scale factor,
\be
\label{5,6}
\rho(t) = \bar{C} \left(a_c(t) + \frac{\gamma}{2}T_c(t)\right)^{-3(\alpha + 1)},
\ee
where $\bar{C}$ is a positive constant. In terms of the commutative variables Eq. (\ref{5}), we have two equivalent ways to write the equations 
that describe the dynamics of the models. In the first one, we write the Einstein's equation for the metric Eq. (\ref{5,5}), use the expression 
for $\rho(t)$ Eq. (\ref{5,6}) and the equation of state for the fluid. In the second way, we introduce the transformations Eq. (\ref{5}) in the 
total Hamiltonian Eq. (\ref{3,5}) and compute the Hamilton's equations for the commutative variables. Since both ways are entirely equivalent, we 
shall use the second way.

We start rewriting the total Hamiltonian $N_{nc} {\mathcal{H}}_{nc}$ Eq. (\ref{3,5}), in terms of the commutative variables Eq. (\ref{5}),
\begin{eqnarray}
\label{6}
N_{nc} {\mathcal{H}}_{nc} & = & -\frac{1}{24} \left( P_{ac} + \frac{\gamma}{2}P_{Tc} \right)^2 
- 6k \left( a_c + \frac{\gamma}{2}T_c \right)^2\nonumber\\
& + & \left(a_c + \frac{\gamma}{2}T_c\right)^{1-3\alpha} \left(P_{Tc} + \frac{\gamma}{2}P_{ac}\right),  
\end{eqnarray}

The Hamilton's equations of motion, obtained using the total Hamiltonian Eq. (\ref{6}) and the
usual PBs among the commutative variables, are,
\ba
\dot a_c &=& \left\{a_c, N_{nc}{\cal{H}}_{nc}\right\} = -\frac{1}{12}\left(P_{ac} + \frac{\gamma}{2}P_{Tc}\right)\nonumber\\ 
&+& \frac{\gamma}{2}\left(a_c + \frac{\gamma}{2}T_c\right)^{1-3\alpha},\label{7}\\
\dot P_{ac} &=& \left\{P_{ac}, N_{nc}{\cal{H}}_{nc}\right\} = 12k\left( a_c + \frac{\gamma}{2}T_c \right)\nonumber\\
&-& (1 - 3\alpha)\left(a_c + \frac{\gamma}{2}T_c\right)^{-3\alpha}\left(P_{Tc} + \frac{\gamma}{2}P_{ac}\right),\label{8}\\
\dot T_c &=& \left\{T_c, N_{nc}{\cal{H}}_{nc}\right\} = -\frac{\gamma}{24}\left( P_{ac} + \frac{\gamma}{2}P_{Tc} \right) \nonumber\\
&+& \left(a_c + \frac{\gamma}{2}T_c\right)^{1-3\alpha},\label{9}\\
\dot P_{Tc} &=& \left\{P_{Tc}, N_{nc}{\cal{H}}_{nc}\right\} = 6\gamma k\left( a_c + \frac{\gamma}{2}T_c\right)\nonumber\\
&-& (1 - 3\alpha)\frac{\gamma}{2}\left(a_c + \frac{\gamma}{2}T_c\right)^{-3\alpha}\left(P_{Tc} + \frac{\gamma}{2}P_{ac}\right)
\label{10}
\ea
Now, we would like to find the NC scale factor behavior (\ref{5,55}). In the general situation, for generic $\alpha$ and $k$, the best we can
do is writing, from Eqs. (\ref{7})-(\ref{10}), a system of two coupled differential equations involving $a_c(t)$, $T_c(t)$
and their time derivatives. This is done in the following way. Combining Eqs. (\ref{8}) and (\ref{10}), we obtain the following
relationship between $P_{Tc}$ and $P_{ac}$,
\be
\label{11}
P_{Tc} = \frac{\gamma}{2}P_{ac} + C,
\ee
where $C$ is an integration constant. Physically, for the commutative case ($\gamma=0$), $C$ represents the fluid energy, which means 
that it is positive. Then, using Eqs. (\ref{7}) and (\ref{9}), we find, to first order in $\gamma$, the following equation
expressing $P_{ac}$ in terms of time derivatives of $a_c$ and $T_c$,
\be
\label{12}
P_{ac} = -12\dot a_c + 6\gamma\dot T_c - \frac{\gamma}{2}C.
\ee
Finally, we introduce the values of $\dot P_{ac}$ Eq. (\ref{8}), $\dot T_c$ Eq. (\ref{9}), $\dot P_{Tc}$ Eq. (\ref{10}), 
$P_{Tc}$ Eq. (\ref{11}) and $P_{ac}$ Eq. (\ref{12}), in the time derivative of Eq. (\ref{7}) and in Eq. (\ref{9}). It gives, to first
order in $\gamma$, the following system of coupled differential equation for $a_c$ and $T_c$,
\begin{eqnarray}
\ddot{a}_{c}(t) & = &  -k ( a_{c}(t) +\frac{\gamma T_{c}(t)}{2})  
- \frac{(1 - 3\alpha)}{2}( \gamma \dot{a}_{c}(t) a_{c}(t)^{-3\alpha}\nonumber \\ 
&-& \frac{C a_{c}(t)^{-3\alpha}}{6} + \frac{C\alpha \gamma T_{c}(t)a_{c}(t)^{-3\alpha - 1}}{4}), \label{13}\\
\dot{T}_{c}(t) & =&  \frac{\gamma \dot{a}_{c}(t)}{2} + a_{c}(t)^{1 - 3\alpha} + \frac{(1 - 3\alpha)\gamma T_{c}(t) a_{c}(t)^{-3\alpha}}{2}. 
\label{14}
\end{eqnarray}
All the information about the noncommutativity is encoded in the parameter $\gamma$. If we set it to zero we recover the usual
commutative model in the gauge $N=a$. In particular, equation (\ref{13}) decouples and we may solve it to obtain the scale factor
dynamics. In order to solve those equations and compute $a_{nc}$ Eq. (\ref{5,55}), we shall have to furnish initial conditions for 
$a_c(t)$, $\dot a_c(t)$ and $T_c(t)$. 
Unfortunately, we cannot find algebraic solutions for $a_c(t)$ and $T_c(t)$, from the system Eqs. (\ref{13})-(\ref{14}), for generic 
values of $k$, $\alpha$, $\gamma$, $C$ and the initial conditions $a_0$, $\dot a_0$ and $T_0$. Where $a_0$, $\dot a_0$ and $T_0$ are, 
respectively, the initial values ($t = 0$) of $a_c(t)$, $\dot a_c(t)$ and $T_c(t)$. Therefore, in what follows, we shall solve that 
system numerically.

\section{The dynamics of the models and the {\it big rip}}
\label{sec:radiation}

Since we shall have to solve the system Eqs. (\ref{13})-(\ref{14}), numerically, we believe that the best way to do that is fixing,
initially, the value of $k$, for each different curvature. Then, for each curvature, we shall investigate how $a_{nc}(t)$ Eq. (\ref{5,55})
behaves for different values of $\alpha$, $\gamma$, $C$ and the initial conditions $a_0$, $\dot a_0$ and $T_0$. 

\subsection{The case k=1}
\label{k=1}

Let us start by fixing $k=1$, it means that the spatial sections have constant positive curvatures. Introducing $k=1$ in the system 
Eqs. (\ref{13})-(\ref{14}), we obtain,

\begin{eqnarray}
\ddot{a}_{c}(t) & = &  - ( a_{c}(t) +\frac{\gamma T_{c}(t)}{2} )  
- \frac{(1 - 3\alpha)}{2}( \gamma \dot{a}_{c}(t) a_{c}(t)^{-3\alpha}  \nonumber \\
&-&\frac{C a_{c}(t)^{-3\alpha}}{6} +\frac{C\alpha \gamma T_{c}(t)a_{c}(t)^{-3\alpha - 1}}{4} ), \label{15}\\
\dot{T}_{c}(t) & =&  \frac{\gamma \dot{a}_{c}(t)}{2} + a_{c}(t)^{1 - 3\alpha} + \frac{(1 - 3\alpha)\gamma T_{c}(t) a_{c}(t)^{-3\alpha}}{2}. 
\label{16}
\end{eqnarray}
Now we are going to solve, numerically, that system for different values of $\alpha$, $\gamma$, $C$ and the initial conditions $a_0$, $\dot{a}_0$ and $T_0$.
It is important to 
mention that the values of those parameters and initial conditions are not entirely arbitrary. There is a constraint between them
given by the Friedmann equation for the initial instant of time. For the present case that constraint is given from Eq. (\ref{a3}),
when it is written in terms of the initial conditions and $k=1$,
\begin{equation}
\label{17}
6\gamma \dot{a}_{0} a_0^{1 - 3\alpha} - 6\dot{a}_0^2 + \frac{C - 12\gamma \dot{a}_0}{a_0^{3\alpha - 1} + 
\frac{(3\alpha - 1)}{2}\gamma T_{0}a_{0}^{3\alpha - 2}} -6(a_0^2 + 2\gamma a_0 T_{0}) = 0.
\end{equation}
Therefore, in order to derive the behavior of $a_{nc}(t)$ Eq. (\ref{5,55}) in terms of the parameters and initial conditions, we
shall vary a given parameter and fix the other parameters and initial conditions with exception of $\dot a_0$. So that, Eq. (\ref{17})
may be satisfied. When we want to vary $\dot{a}_0$, we shall fix all other parameters and initial conditions with exception of $C$.
So that, Eq. (\ref{17}) may be satisfied. The choices of $\dot{a}_0$ and $C$ as the quantities to be left free so that Eq. (\ref{17})
may be satisfied, are arbitrary and do not modify our conclusions.

After solving, numerically, the system Eqs. (\ref{15})-(\ref{16}), for many different
values of all parameters and initial conditions, we reach the following conclusions.
The general behavior of $a_{nc}(t)$ Eq. (\ref{5,55}) describes a universe that starts
to expand in an accelerated rate from its initial size $a_0$ at $t=0$, and ends, after
a finite time interval ($t_{br}$), in a {\it big rip} singularity. That general
behavior of $a_{nc}(t)$ is qualitatively similar to the corresponding commutative
scale factor, the differences being of quantitative nature. Let us see, now, the
specific properties of $a_{nc}(t)$ due to each parameter and initial condition.

\subsubsection{Varying $\alpha$}

We start computing $a_c(t)$ and $T_c(t)$ and eventually the physical scale factor $a_{nc}(t)$ Eq. (\ref{5,55}) from the system
Eqs. (\ref{15})-(\ref{16}), by varying $\alpha < -1$ and fixing all other parameters and initial conditions. 
For models with different values of $\alpha$, we notice that: the more negative $\alpha$, the more quickly the NC scale
factor reaches the {\it big rip} singularity. Therefore, the more repulsive the fluid, the more quickly $a_{nc}(t)$ reaches the
{\it big rip} singularity. In fact, that conclusion agrees with the state equation of the fluid: $p = \alpha\rho$ and with the
commutative model. As an example of that conclusion, we can see Figure 1.

\begin{figure}[!htb]
	\centering
	\begin{minipage}[c]{0.49\linewidth}
		\centering
		\includegraphics[width=5cm]{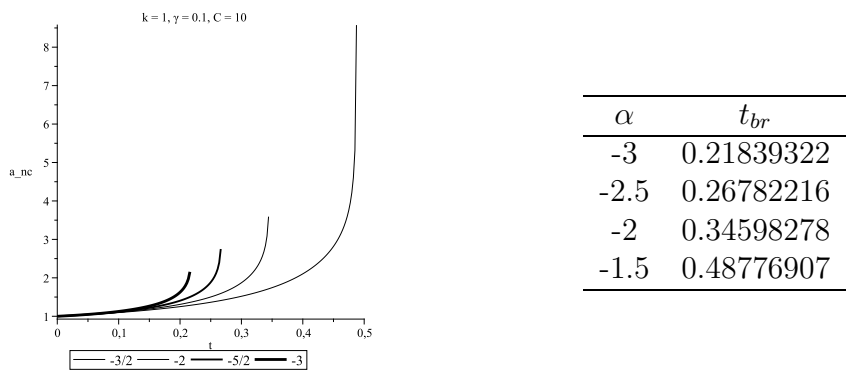}
	\end{minipage}
	\hfill
	\begin{minipage}[c]{0.49\linewidth}
		\centering
		\begin{tabular}{ccc}\hline
			$\alpha$ & $t_{br}$ \\ \hline
			-3 & 0.21839322 \\
			-2.5 & 0.26782216 \\
			-2 & 0.34598278 \\
			-1.5 & 0.48776907 \\ \hline
		\end{tabular}
	\end{minipage}
		\caption{NC scale factor as a function of $t$, for $k=1$, $\gamma = 0.1$, $C = 10$, $a_0 = 1$ and $T_0 = 0$. 
		Particularly, for that choice of parameters and initial conditions we obtain for the four values of $\alpha$ the same initial condition $\dot{a}_0 = 0.786$.
		The table shows the amount of time $a_{nc}(t)$ takes to reach the {\it big rip} singularity after start expanding at $t=0$, 
		for each different value of $\alpha$.}\label{fig1}
\end{figure}

\subsubsection{Varying $\gamma$}

After solving, numerically, the system Eqs. (\ref{15})-(\ref{16}), for many different values of $\gamma$, the NC parameter, 
keeping fix all other parameters and initial conditions, we reach the following conclusions. The {\it big rip} singularity 
cannot be avoided, due to the noncommutativity. In fact, whenever we increase the modulus of $\gamma$, positive or negative,
the time it takes, for the NC scale factor to reach the {\it big rip} singularity, diminishes. In this way, noncommutativity 
behaves as an additional repulsive force to the one already produced by the phantom fluid, helping the accelerated expansion 
of the universe. It means that, for any noncommutative model $t_{br}$ will be always greater than the corresponding time in the
commutative model. As an example of that conclusion, we can see Figure 2.

\begin{figure}[!htb]
	\centering
	\begin{minipage}[c]{0.49\linewidth}
		\centering
		\includegraphics[width=5cm]{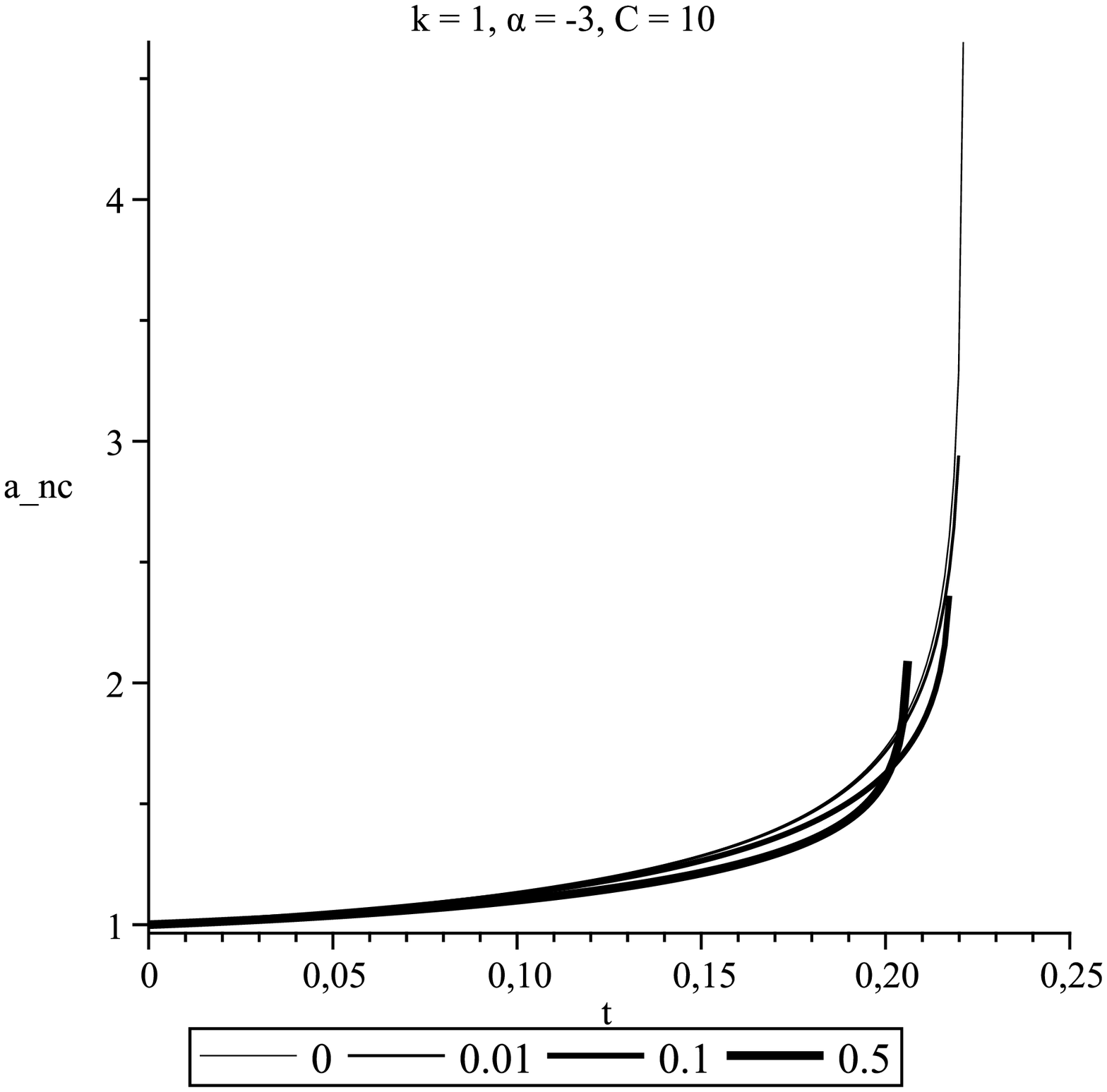}
	\end{minipage}
	\hfill
	\begin{minipage}[c]{0.49\linewidth}
		\centering
		\begin{tabular}{ccc}\hline
			$\gamma$ & $t_{br}$ & $\dot{a}_0$ \\ \hline
			0.5 & 0.20666878 & 0.604 \\
			0.1 & 0.21839322 & 0.768 \\
			0.01 & 0.22106239 & 0.812 \\
			0 & 0.22151935 & 0.816 \\ \hline
		\end{tabular}
	\end{minipage}
		\caption{NC scale factor as a function of $t$, for $k = 1$, $\alpha = -3$, $C = 10$, $a_0 = 1$ and $T_0 = 0$.
		The table shows the amount of time $a_{nc}(t)$ takes to reach the {\it big rip} singularity after start expanding at $t=0$, 
		for each different value of $\gamma$.}\label{fig2}
\end{figure}

Although qualitatively both positive and negative values of $\gamma$ behave as repulsive forces, quantitatively they have different
strengths. In order to study this property, we consider two different models where in the first model $\gamma$ is positive and in the 
second $\gamma$ is negative, but has the same modulus than in the first. Apart from that, all the other parameters and initial conditions 
have the same values in both models. After studying many different models of that type, we conclude that the NC scale factor in the models 
with $\gamma < 0$ go to the {\it big rip} singularity quicker than in the models with $\gamma > 0$. Therefore, the strength of the repulsive
force for $\gamma < 0$ is greater than for $\gamma > 0$. As an example of that conclusion, we can see Figure 3.

\begin{figure}[!htb]
	\centering
	\begin{minipage}[c]{0.49\linewidth}
		\centering
		\includegraphics[width=5cm]{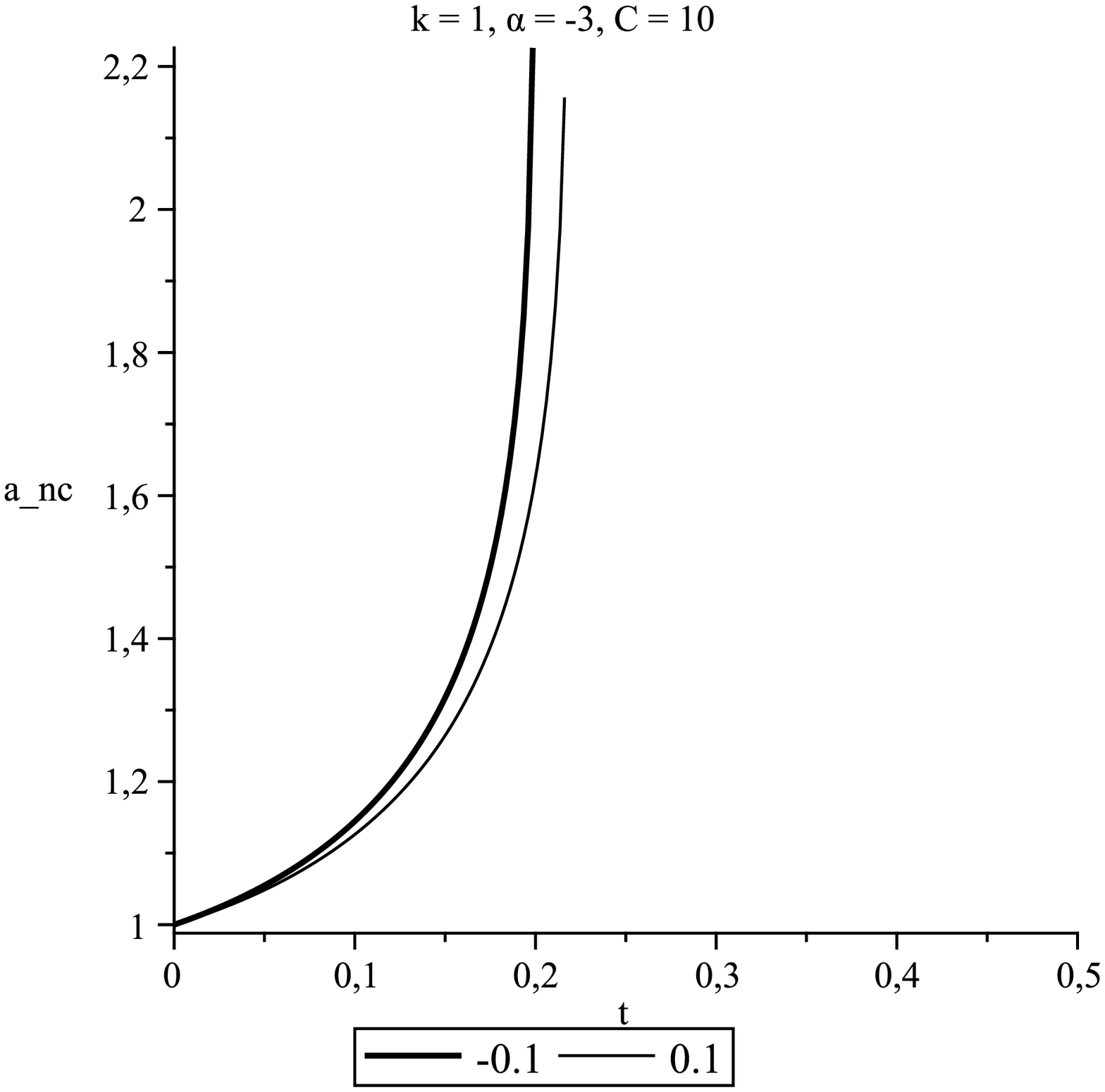}
	\end{minipage}
	\hfill
	\begin{minipage}[c]{0.49\linewidth}
		\centering
		\begin{tabular}{ccc}\hline
			$\gamma$ & $t_{br}$ & $\dot{a}_0$ \\ \hline
			-0.1 &  0.20003181 & 0.868 \\
			0.1 & 0.21839322 & 0.768  \\ \hline
		\end{tabular}
	\end{minipage}
	\caption{NC scale factor as a function of $t$, for $k = 1$, $\alpha = -3$, $C = 10$, $a_0 = 1$ and $T_0 = 0$.
	The table shows the amount of time $a_{nc}(t)$ takes to reach the {\it big rip} singularity after start expanding at $t=0$, 
		for each different value of $\gamma$.}\label{fig3}
\end{figure}

\subsubsection{Varying $C$}

After solving, numerically, the system Eqs. (\ref{15})-(\ref{16}), for many different values of $C$, the parameter associated to 
the fluid energy, keeping fix all other parameters and initial conditions, we reach the following conclusions. If one increases 
the value of $C$, the NC scale factor goes quicker to the {\it big rip} singularity. In other words, if one increases the fluid
energy it becomes more repulsive and expands more rapidly. That result agrees with the corresponding one in the commutative model. 
As an example of that conclusion, we can see Figure 4.

\begin{figure}[!htb]
	\centering
	\begin{minipage}[c]{0.49\linewidth}
		\centering
		\includegraphics[width=5cm]{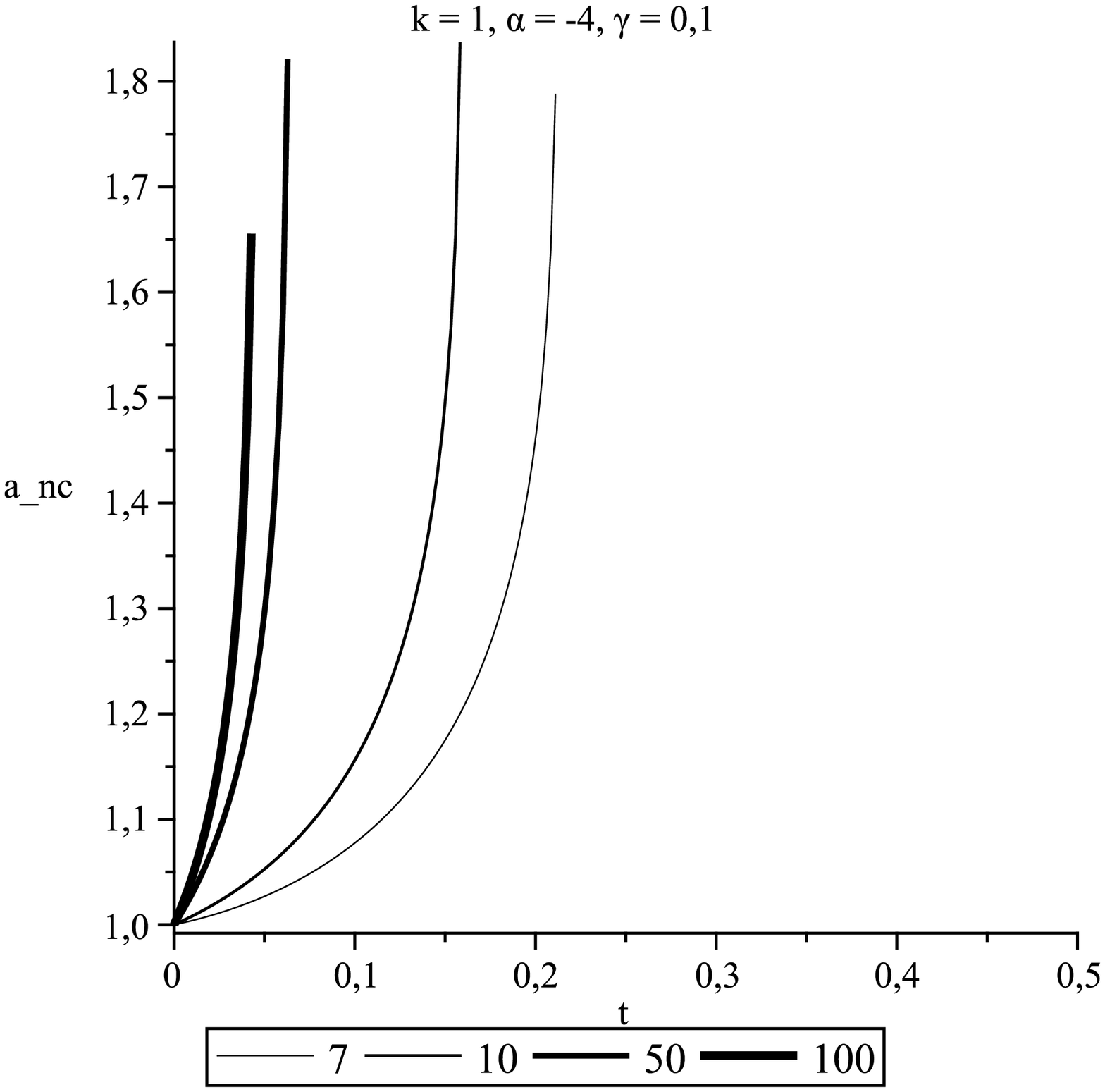}
	\end{minipage}
	\hfill
	\begin{minipage}[c]{0.49\linewidth}
		\centering
		\begin{tabular}{ccc}\hline
			$C$ & $t_{br}$ & $\dot{a}_0$ \\ \hline
			100 & 0.04475916 & 3.908 \\
			50 & 0.06390654 & 2.658 \\
			10 & 0.15945706 & 0.768\\
			7 &  0.21266708 & 0.361 \\ \hline
		\end{tabular}
	\end{minipage}
	\caption{NC scale factor as a function of $t$, for $k = 1$, $\alpha = -4$, $\gamma = 0.1$, $a_0 = 1$ and $T_0 = 0$.
	The table shows the amount of time $a_{nc}(t)$ takes to reach the {\it big rip} singularity after start expanding at $t=0$, 
		for each different value of $C$.}\label{fig4}
\end{figure}
Let us see, now, how the dynamics of the NC models, with $k = 1$, depend on the initial conditions.

\subsubsection{Varying $a_0$}
\label{$a_0$}

After solving, numerically, the system Eqs. (\ref{15})-(\ref{16}), for many different values of $a_0$, the initial value associated to 
the scale factor, keeping fix all other parameters and initial conditions, we reach the following conclusions. The greater the value of 
$a_0$, the more quickly the NC scale factor reaches the {\it big rip} singularity. Therefore, universes that start with greater values of
$a_0$ will end quicker. That result agrees with the corresponding one in the commutative model. As an example of that conclusion, we can see Figure 5.

\begin{figure}[!htb]
	\centering
	\begin{minipage}[c]{0.49\linewidth}
		\centering
		\includegraphics[width=5cm]{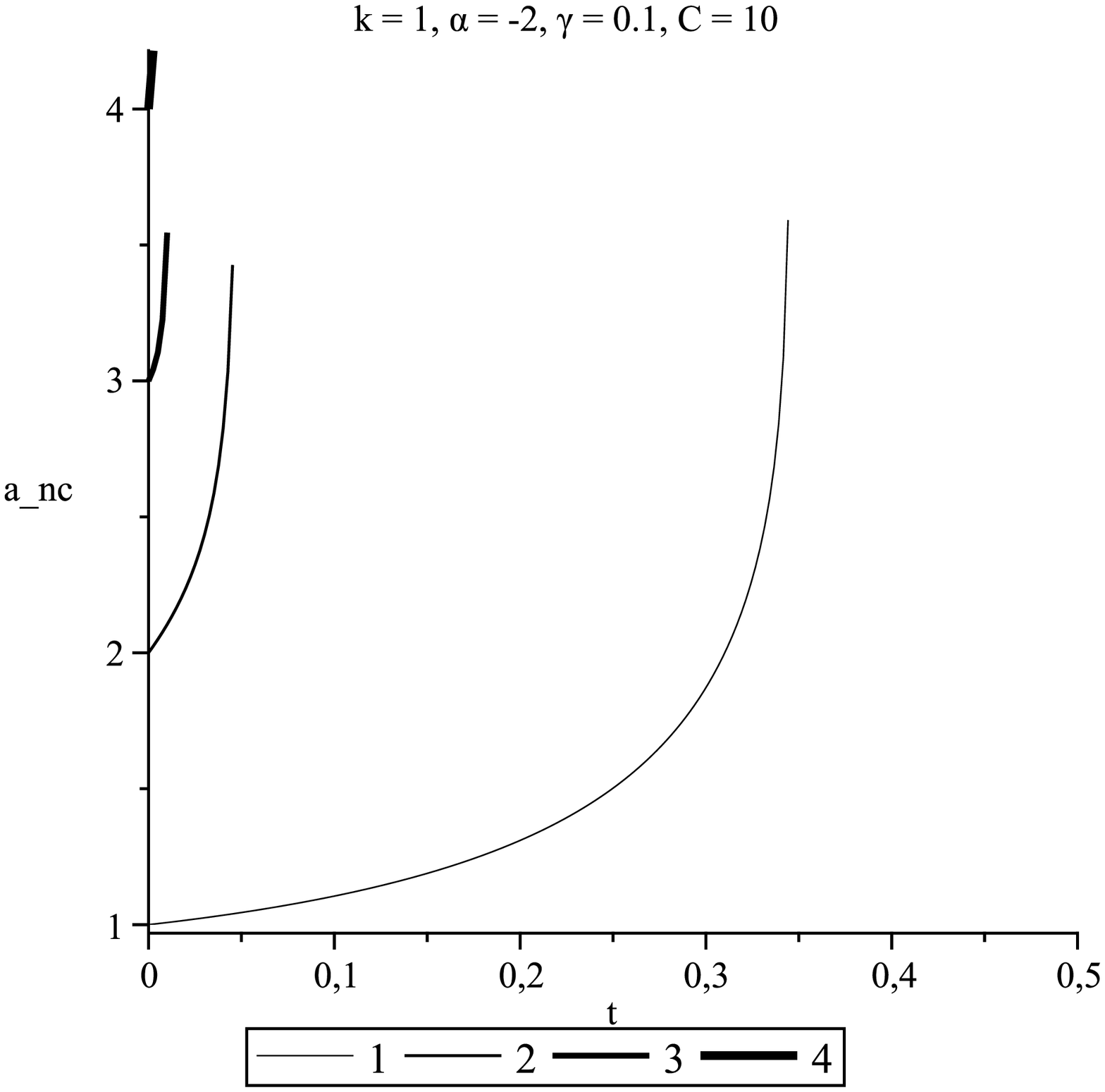}
	\end{minipage}
	\hfill
	\begin{minipage}[c]{0.49\linewidth}
		\centering
		\begin{tabular}{ccc}\hline
			$a_0$ & $t_{br}$ & $\dot{a}_0$\\ \hline
			4 & 0.00354608 & 16.491 \\
			3 & 0.01222068 & 15.524 \\
			2 & 0.04757719 & 9.421 \\
			1 & 0.34598278 & 0.768 \\ \hline
		\end{tabular}
	\end{minipage}
	\caption{NC scale factor as a function of $t$, for $k = 1$, $\alpha = -2$, $\gamma = 0.1$, $C = 10$ and $T_0 = 0$.
	The table shows the amount of time $a_{nc}(t)$ takes to reach the {\it big rip} singularity after start expanding at $t=0$, 
		for each different value of $a_0$.}\label{fig5}
\end{figure}

\subsubsection{Varying $\dot{a}_0$}

After solving, numerically, the system Eqs. (\ref{15})-(\ref{16}), for many different values of $\dot{a}_0$, the initial value associated to 
the scale factor velocity, keeping fix all other parameters and initial conditions, we reach the following conclusions. The greater the value of 
$\dot{a}_0$, the more quickly the NC scale factor reaches the {\it big rip} singularity. This result was expected since, if one increases the initial
$a_0$ velocity, $a_0$ will expand quicker. That result agrees with the corresponding one in the commutative model. As an example of that conclusion, we can see Figure 6.

\begin{figure}[!htb]
	\centering
	\begin{minipage}[c]{0.49\linewidth}
		\centering
		\includegraphics[width=5cm]{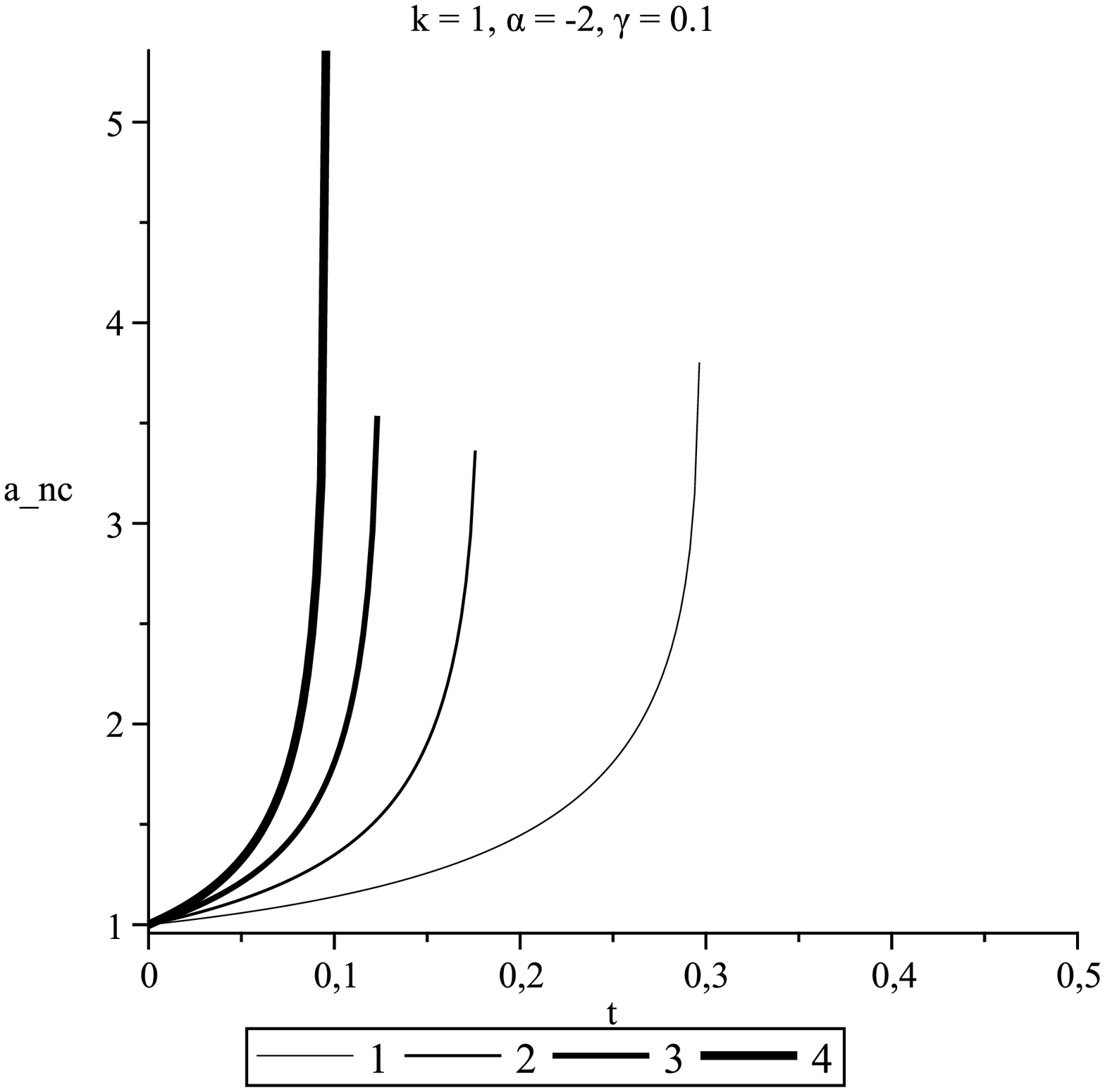}
	\end{minipage}
	\hfill
	\begin{minipage}[c]{0.49\linewidth}
		\centering
		\begin{tabular}{ccc}\hline
			$\dot{a}_0$ & $t_{br}$ & $C$ \\ \hline
			4 & 0.09563811 & 104.400 \\
			3 & 0.12491159 & 61.800 \\
			2 & 0.17836852 & 31.200 \\
			1 & 0.29773577 & 12.600 \\ \hline
		\end{tabular}
	\end{minipage}
	\caption{NC scale factor as a function of $t$, for $k = 1$, $\alpha = -2$, $\gamma = 0.1$, $a_0 = 1$ and $T_0 = 0$.
	The table shows the amount of time $a_{nc}(t)$ takes to reach the {\it big rip} singularity after start expanding at $t=0$, 
		for each different value of $\dot{a}_0$.}\label{fig6}
\end{figure}

\subsubsection{Varying $T_0$}

After solving, numerically, the system Eqs. (\ref{15})-(\ref{16}), for many different values of $T_0$, the initial value associated to 
the variable $T$, keeping fix all other parameters and initial conditions, we reach the following conclusions. Here, we obtain two
different results depending whether $\gamma$ is positive or negative. For $\gamma > 0$, the greater the value of $T_0$, the more quickly 
the NC scale factor reaches the {\it big rip} singularity. This result is similar to what happened when we varied the commutative
scale factor initial value ($a_0$), in Subsubsection \ref{$a_0$}. This happens because $a_{nc}$ Eq. (\ref{5,55}), is a crescent linear function of $a_0$ and $T_0$
for $\gamma > 0$. Therefore, increasing $T_0$ the initial value of $a_{nc}$ also increases, in the same way that happened when $a_0$ was
increased. As an example of that conclusion, we can see Figure 7.

\begin{figure}[!htb]
	\centering
	\begin{minipage}[c]{0.49\linewidth}
		\centering
		\includegraphics[width=5cm]{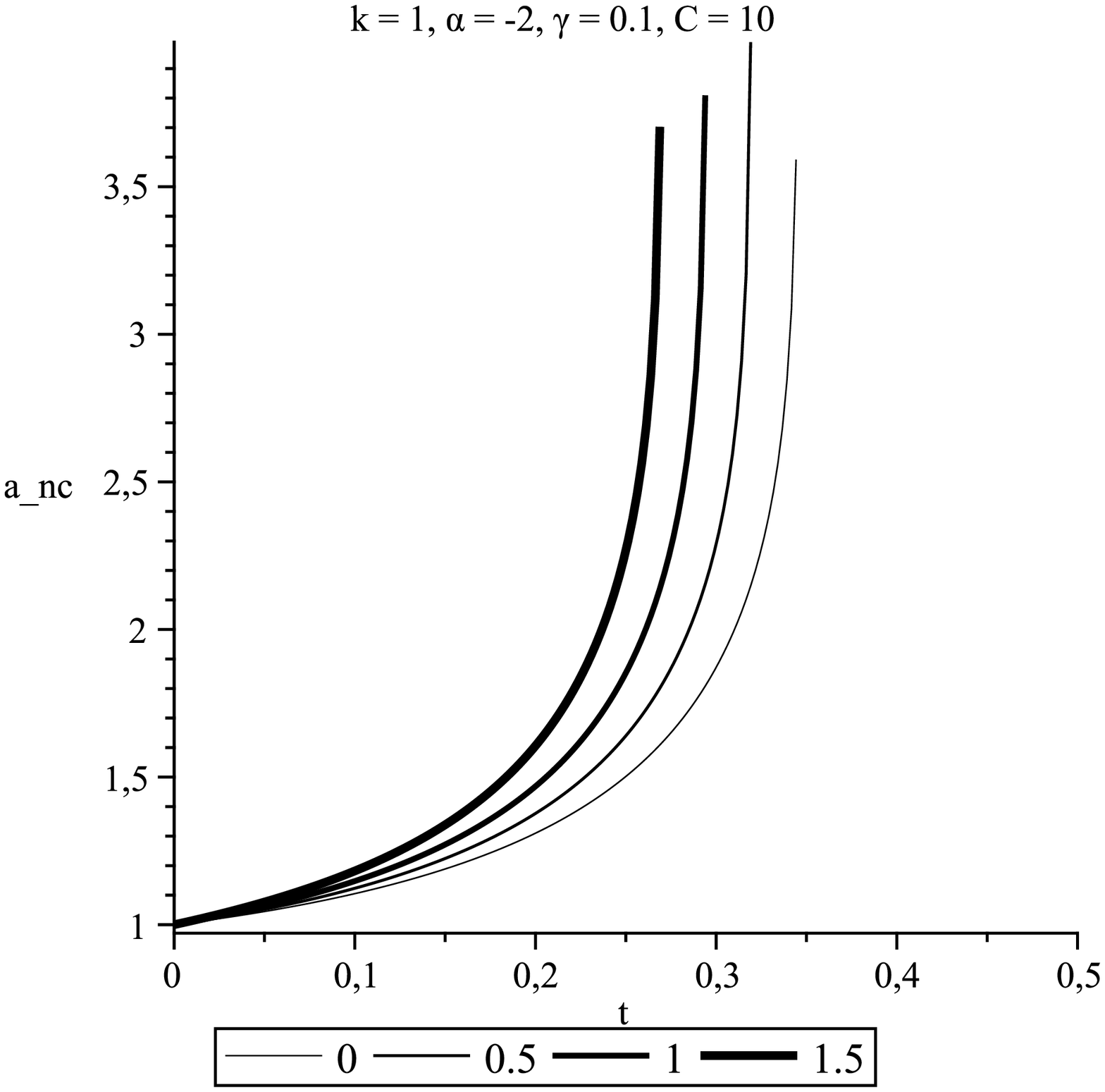}
	\end{minipage}
	\hfill
	\begin{minipage}[c]{0.49\linewidth}
		\centering
		\begin{tabular}{ccc}\hline
			$T_0$ & $t_{br}$ & $\dot{a}_0$ \\ \hline
			1,5 & 0.27031174 & 1.334 \\
			1 & 0.29520983 & 1.069 \\
			0,5 & 0.32003741 & 0.891 \\
			0 & 0.34598278 & 0.768 \\ \hline
		\end{tabular}
	\end{minipage}
	\caption{\footnotesize NC scale factor as a function of $t$, for $k = 1$, $\alpha = -2$, $\gamma = 0.1$, $C = 10$ and $a_0 = 1$.
	The table shows the amount of time $a_{nc}(t)$ takes to reach the {\it big rip} singularity after start expanding at $t=0$, 
		for each different value of $T_0$.}\label{fig7}
\end{figure}

For $\gamma < 0$, the opposite result happens. The greater the value of $T_0$, the more slowly 
the NC scale factor reaches the {\it big rip} singularity. This happens because, now, increasing $T_0$ the initial value of $a_{nc}$ decreases.
As an example of that conclusion, we can see Figure 8. Since the commutative scale factor does not depend on $T$, the above results have no correspondent ones in the commutative model.

\begin{figure}[!htb]
	\centering
	\begin{minipage}[c]{0.49\linewidth}
		\centering
		\includegraphics[width=5cm]{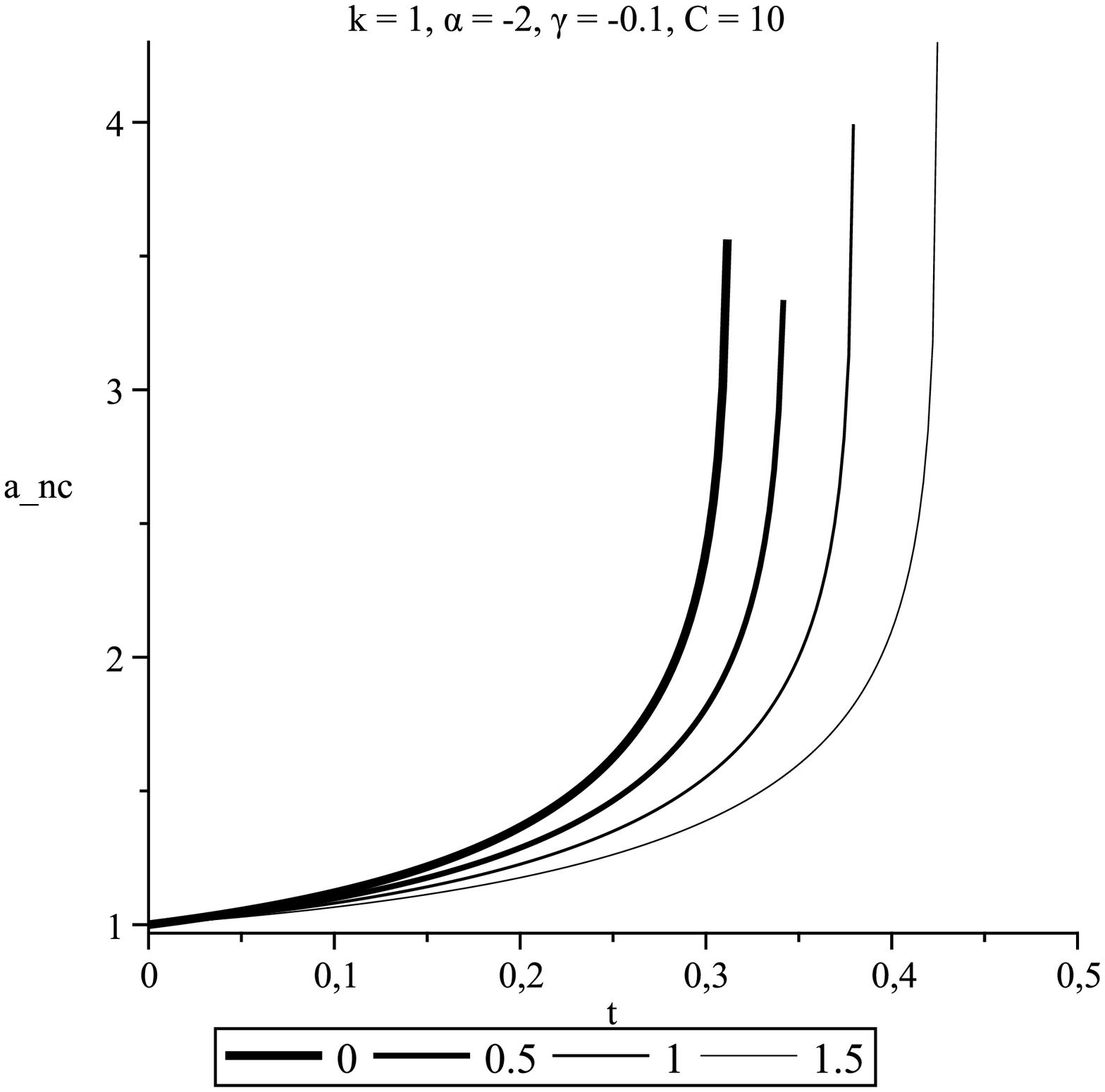}
	\end{minipage}
	\hfill
	\begin{minipage}[c]{0.49\linewidth}
		\centering
		\begin{tabular}{ccc}\hline
			$T_0$ & $t_{br}$ & $\dot{a}_0$ \\ \hline
			0 & 0.31312873 & 0.868 \\		
			0.5 & 0.34400296 & 0.720 \\
			1 & 0.38020472 & 0.603 \\
			1.5 & 0.42514472 & 0.509 \\ \hline
		\end{tabular}
	\end{minipage}
	\caption{\footnotesize NC scale factor as a function of $t$, for $k = 1$, $\alpha = -2$, $\gamma = -0.1$, $C = 10$ and $a_0 = 1$.
	The table shows the amount of time $a_{nc}(t)$ takes to reach the {\it big rip} singularity after start expanding at $t=0$, 
		for each different value of $T_0$.}\label{fig8}
\end{figure}

\subsection{The cases $k = 0$ and $k = -1$}
\label{k=0,-1}

Here, we must proceed in the same way we did in the last subsection. First, we rewrite the system Eqs. (\ref{13})-(\ref{14}), for the cases $k=0$ and $k=-1$.
Then, we solve the resulting system, numerically, in order to investigate how $a_{nc}(t)$ Eq. (\ref{5,55}) behaves for different values of $\alpha$, $\gamma$, 
$C$ and the initial conditions $a_0$, $\dot a_0$ and $T_0$. As in the case $k=1$, the values of those parameters and initial conditions are not entirely arbitrary, 
they are constrained by the Friedmann equation for the initial instant of time. For the present cases those two constraints are given from Eq. (\ref{a3}),
when it is written in terms of the initial conditions and $k=0$ and the initial conditions and $k=-1$.

We solved, numerically, the appropriated systems obtained from Eqs. (\ref{13})-(\ref{14}), for the cases $k=0$ and $k=-1$, and investigated how $a_{nc}(t)$ 
Eq. (\ref{5,55}) behaves for different values of all the parameters and initial conditions. After computing a great number of solutions for different values
of parameters and initial conditions, for both cases $k=0$ and $k=-1$, we reach the following conclusions. In the same way as in the previous case, 
the general behavior of $a_{nc}(t)$ Eq. (\ref{5,55}), for both cases, describe universes that start
expanding, in an accelerated rate, from the initial size $a_0$ at $t=0$, and end, after
a finite time interval ($t_{br}$), in {\it big rip} singularities. That general
behavior of $a_{nc}(t)$, for both cases, are qualitatively similar to the corresponding commutative
scale factors, the differences being of quantitative nature.
We also found that, for both cases $k=0$ and $k=-1$,
the {\it big rip} singularities cannot be avoided, due to noncommutativity, in the present cases.
On the other hand, there are quantitative differences in the behavior of $a_{nc}(t)$, between the three cases. Let us
present those similarities and differences for each parameter and initial condition.

\subsubsection{$\alpha$}
As in the case $k=1$, here, for $k=0$ and $k=-1$, the more negative $\alpha$, the more quickly the NC scale factor reaches the {\it big rip} singularity. Therefore, the more repulsive the fluid, the more quickly $a_{nc}(t)$ reaches the {\it big rip} singularity. That result agrees with the corresponding ones in the commutative models, for $k=0$ and $k=-1$. 
Examples of these cases, for $k=0$ and $k=-1$, would produce figures, qualitatively, very similar to Figure 1, for the case $k=1$.

\subsubsection{$\gamma$}
\label{gamma}

As in the case $k=1$, here, for $k=0$ and $k=-1$, whenever we increase the modulus of $\gamma$, positive or negative, the NC scale factor goes quicker to the {\it big rip} singularity. In this way, the noncommutativity behaves as an additional repulsive force to the phantom fluid, helping the accelerated expansion of the universe.
It means that, for any noncommutative model, with $k=0$ or $k=-1$, $t_{br}$ will be always great than the corresponding times
in the commutative models.
After studying many different models of that type, we conclude that $a_{nc}(t)$ in the models 
with $\gamma < 0$ goes to the {\it big rip} singularity quicker than in the models with $\gamma > 0$. Therefore, the strength of the repulsive force for $\gamma < 0$ is greater than for $\gamma > 0$. Examples of these cases, for $k=0$ and $k=-1$, would produce figures, qualitatively, very similar to Figures 2 and 3, for the case $k=1$.

\subsubsection{$C$}
As in the case $k=1$, here, for $k=0$ and $k=-1$, if one increases the value of $C$, the NC scale factor goes quicker to the {\it big rip} singularity. In other words, if one increases the fluid energy it becomes more repulsive and expands more rapidly. That result agrees with the corresponding ones in the commutative models, for $k=0$ and $k=-1$.
Examples of these cases, for $k=0$ and $k=-1$, would produce figures, qualitatively, very similar to Figure 4, for the case $k=1$.

\subsubsection{$a_0$}
\label{a0}

As in the case $k=1$, here, for $k=0$ and $k=-1$, the greater the value of 
$a_0$, the more quickly the NC scale factor reaches the {\it big rip} singularity. Therefore, universes that start with greater values of $a_0$ will end quicker. That result agrees with the corresponding ones in the commutative models, for $k=0$ and $k=-1$.
Examples of these cases, for $k=0$ and $k=-1$, would produce figures, qualitatively, very similar to Figure 5, for the case $k=1$.

\subsubsection{$\dot{a}_0$}
\label{dota0}

As in the case $k=1$, here, for $k=0$ and $k=-1$, the greater the value of 
$\dot{a}_0$, the more quickly the NC scale factor reaches the {\it big rip} singularity. This result was expected since, if one increases the initial $a_0$ velocity, $a_0$ will expand quicker. That result agrees with the corresponding ones in the commutative models, for $k=0$ and $k=-1$.
Examples of these cases, for $k=0$ and $k=-1$, would produce figures, qualitatively, very similar to Figure 6, for the case $k=1$.

\subsubsection{$T_0$}
As in the case $k=1$, here, for $k=0$ and $k=-1$, for $\gamma > 0$, the greater the value of $T_0$, the more quickly 
the NC scale factor reaches the {\it big rip} singularity.
For $\gamma < 0$, the opposite result happens. The greater the value of $T_0$, the more slowly 
$a_{nc}(t)$ reaches the {\it big rip} singularity. Examples of these cases, for $k=0$ and $k=-1$, would produce figures, qualitatively, very similar to Figures 7 and 8, for the case $k=1$. Since the commutative scale factors, for models with $k=0$ or $k=-1$, do not depend on $T$, the above results have no correspondent ones in the commutative models.

\subsection{Comparison between different values of $k$}
\label{ks}

After studying the behavior of $a_{nc}(t)$ Eq. (\ref{5,55}), for different values of $k$, we noticed that although it
behaves qualitatively in a very similar way in all three cases, it presents some quantitative differences depending on 
the value of $k$. More precisely, if we fix all parameters and initial conditions with the exception of $k$, we observe
that the NC scale factor reaches the {\it big rip} singularity firstly for the model with $k=-1$, secondly for the
model with $k=0$ and lastly for the model with $k=1$. That result agrees with the corresponding one in the commutative models.
As an example of that behavior, we can see Figure 9.

\begin{figure}[!htb]
	\centering
	\begin{minipage}[c]{0.49\linewidth}
		\centering
		\includegraphics[width=5cm]{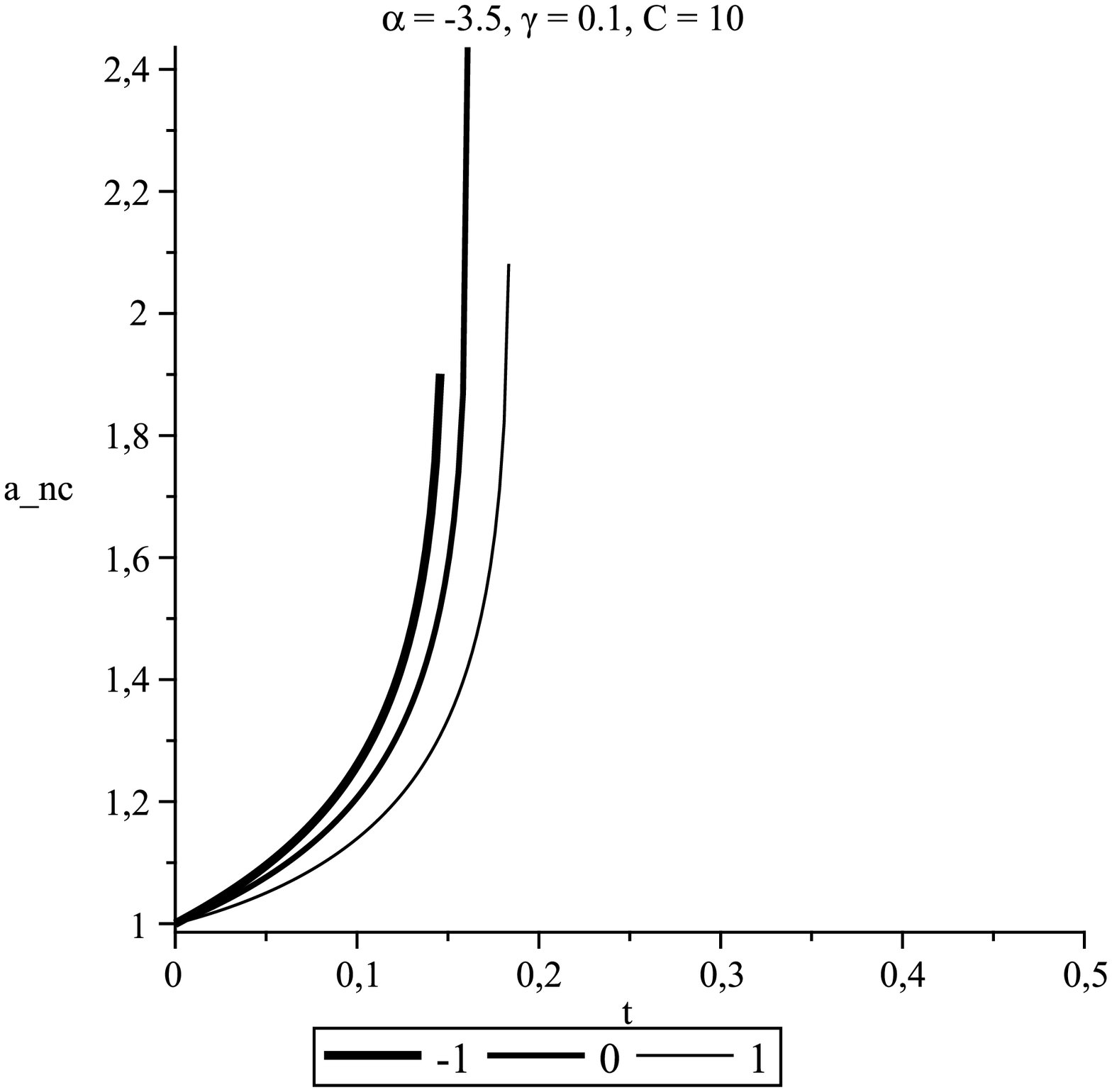}
	\end{minipage}
	\hfill
	\begin{minipage}[c]{0.49\linewidth}
		\centering
		\begin{tabular}{ccc}\hline
			$k$ & $t_{br}$ & $\dot{a}_0$ \\ \hline
			-1 & 0.14801374 & 1.584\\
			0 & 0.16098371 & 1.242 \\
			1 & 0.18433758 & 0.768 \\ \hline
		\end{tabular}
	\end{minipage}
	\caption{\footnotesize NC scale factor as a function of $t$, for $\alpha = -3.5$, $\gamma = 0.1$, $C = 10$, 
	$a_0 = 1$ and $T_0 = 0$.
	The table shows the amount of time $a_{nc}(t)$ takes to reach the {\it big rip} singularity after start expanding at $t=0$, 
		for each different value of $k$.}\label{fig9}
\end{figure}

\section{Estimates for $\gamma$ and the time intervals till the end of the Universe}
\label{sec:endofuniverse}

In the present section, we want to give some estimates for the NC parameter $\gamma$. 
Then, using those estimated values of $\gamma$, we shall compute the corresponding time intervals till the end 
of the Universe ({\it big rip}). 

In order to obtain estimates for $\gamma$, let us start computing $\dot{a}_{nc}$, from the total NC 
Hamiltonian Eq. (\ref{3,5}) in the gauge $N_{nc}=a_{nc}$,
\begin{equation}
\label{18}
\dot{a}_{nc} = \left\{ a_{nc}, N_{nc}H_{nc}  \right\} = - \frac{1}{12}P_{a_{nc}} + \frac{\gamma}{a_{nc}^{3\alpha - 1}}.
\end{equation}
From the above equation (\ref{18}), we may compute the value of $P_{a_{nc}}$ as,
\begin{equation}
\label{19}
P_{a_{nc}} = -12\left(\dot{a}_{nc} - \frac{\gamma}{a_{nc}^{3\alpha - 1}}\right).
\end{equation}
Now, using again $N_{nc}H_{nc}$ Eq. (\ref{3,5}), we obtain $\dot{P}_{T_{nc}}$,
\begin{equation}
\label{20}
\dot{P}_{T_{nc}} = \left\{ P_{T_{nc}}, N_{nc}H_{nc}  \right\} = -\gamma P_{T_{nc}} (1 - 3\alpha)a^{-3\alpha}+12ka_{nc}.
\end{equation}
As a simplification, we shall consider that, from its birth until the time the present accelerated
expansion started, the Universe was dominated by a dust perfect fluid ($\alpha=0$). We shall, also consider, that the Universe
has flat spatial sections ($k=0$). Therefore, under those conditions Eq. (\ref{20}) is simplified to,
\begin{equation}
\label{21}
\dot{P}_{T_{nc}} = -\gamma P_{T_{nc}}.
\end{equation}
That equation may be easily integrated to give,
\begin{equation}
\label{22}
P_{T_{nc}} = P_{T_{nc0}}e^{-\gamma t},
\end{equation}
where $P_{T_{nc0}}$ is the initial value of the momentum canonically conjugated to $T_{nc}$.
We may, now, introduce Eqs. (\ref{19}) and (\ref{22}) in the Friedmann equation, which is
obtained by setting $N_{nc}H_{nc}$ Eq. (\ref{3,5}) equal to zero. If we take in account the simplified
conditions ($\alpha=k=0$), we find, to first order in $\gamma$, the following Friedmann equation,
\begin{equation}
\label{23}
-6\dot{a}^2_{nc} + 12 \dot{a}_{nc}\gamma a_{nc} + P_{T_{nc0}}e^{-\gamma t} a_{nc} = 0 .
\end{equation}
Observing Eq. (\ref{23}), we see that for very large time intervals the last term in the LHS can be
neglected in comparison with the other terms, due to the exponential function. Therefore, imposing that
additional condition we may solve the resulting Friedmann equation and find the following expression
for $a_{nc}(t)$,
\begin{equation}
\label{24}
a_{nc}(t) = a_{0} e^{2\gamma t}.
\end{equation}
We shall estimate the value of $\gamma$ from that equation. In order to do that, we must give
$a_0$ which is the initial scale factor value. We cannot choose $a_0=0$ because it would be impossible
to obtain any value for $\gamma$. Therefore, we shall choose it as close to zero as we can. Under our 
present computational conditions it is $a_0=10^{-40}$. We must also give the values of time ($t_h$)
and scale factor ($a_h$), for the beginning of the present accelerated expansion of the Universe.
As an example, in Table \ref{comparison}, we computed ten values of $\gamma$ using ten different values 
of $a_h$ and $t_h$. We obtained those values considering that the present mass density parameter ($\Omega_{m0}$) 
is equal to 0.3 and the present Hubble constant ($H_0$) is equal to 70 (km/s)/Mpc. From Table \ref{comparison},
we observe that $\gamma$ increases as the initial time of the present accelerated expansion of the Universe $t_h$ 
approaches the initial moments of the Universe. That result is expected since noncommutativity should had been
more important at the beginning of the Universe.

\begin{table}[h!]
\caption{{\protect\footnotesize {A table
with $10$ different values of $\gamma$ and the corresponding time interval till the big rip $t_{br}$.
}}}
\centering
{\scriptsize\begin{tabular}{|c|c|c|c|c|}
\hline $a_h$ & $t_h(Gyear)$ & $\gamma$ & $\dot{a}_h$ & $t_{br}(Gyear)$\\ \hline
$1$ & $13.4560$ & $1.085233269\times10^{-16}$ & $3.318464311\times10^{-20}$ & $1.27476810629122$\\ \hline
$0.9$ & $12.0224$ & $1.213251436\times10^{-16}$ & $2.968108672\times10^{-20}$ & $1.53305821917808$\\ \hline
$0.8$ & $10.5167$ & $1.385179732\times10^{-16}$ & $2.599480917\times10^{-20}$ & $1.87174613140538$\\ \hline
$0.7$ & $8.9511$ & $1.625090595\times10^{-16}$ & $2.215472070\times10^{-20}$ & $2.32854826230340$\\ \hline
$0.6$ & $7.3488$ & $1.976092422\times10^{-16}$ & $1.821680467\times10^{-20}$ & $2.96862186073059$\\ \hline
$0.5$ & $5.7470$ & $2.521837672\times10^{-16}$ & $1.427166297\times10^{-20}$ & $3.91549720953831$\\ \hline
$0.4$ & $4.1973$ & $3.444505259\times10^{-16}$ & $1.044570358\times10^{-20}$ & $5.43375634195840$\\ \hline
$0.3$ & $2.7629$ & $5.216261982\times10^{-16}$ & $6.894448455\times10^{-21}$ & $8.19626490360223$\\ \hline
$0.2$ & $1.5148$ & $9.471695377\times10^{-16}$ & $3.793376470\times10^{-21}$ & $14.4565829528158$\\ \hline
$0.1$ & $0.5370$ & $2.651364319\times10^{-15}$ & $1.351320304\times10^{-21}$ & $37.4975646879756$\\ \hline
\end{tabular}
}
\label{comparison}
\end{table}

Now, for a given $\gamma$, we want to compute the corresponding time interval till the end of the Universe ({\it big rip}).
It means that, the Universe is no longer dominated by dust. It is dominated, now, by a phantom perfect fluid.
Therefore, we must take the given value of $\gamma$ and solve the corresponding system Eqs. (\ref{13})-(\ref{14}),
for $k=0$. As an example, in Table \ref{comparison}, we computed ten values of the time 
interval till the {\it big rip} ($t_{br}$), using the ten different values of $\gamma$ and $a_h$, already mentioned in 
Table \ref{comparison}. $a_h$ represents, now, the initial scale factor. We choose a phantom perfect fluid with 
$\alpha=-1.01$, which is compatible with present observations \cite{riess}. The perfect fluid energy density is
given by $C=6\Omega_{de}H^2_0$, where we took the dark energy mass parameter $\Omega_{de}$ to be equal to 0.7.
The initial scale factor velocity ($\dot{a}_h$), given in Table \ref{comparison} were computed with the help of the 
appropriated Friedmann equations, obtained from Eq. (\ref{a3}), for the values of the parameters and initial values 
already given. From Table \ref{comparison}, we observe that $t_{br}$ increases when $\gamma$ increases, which seems contradictory
to the results derived in Subsubsection \ref{gamma}. On the other hand, we notice, also from Table \ref{comparison}, that
when $\gamma$ increases, both $a_h$ and $\dot{a}_h$ decrease. Therefore, from the results of Subsubsections \ref{a0} and
\ref{dota0}, we understand that $t_{br}$ increases not because $\gamma$ increases but because both $a_h$ and $\dot{a}_h$ 
decrease. In fact, the values of $a_h$ are much bigger than the values of $\gamma$, therefore $a_h$ must influence the
behavior of $t_{br}$ more strongly than $\gamma$.

\section{Conclusions}
\label{sec:conclusions}

We conclude that noncommutativity modifies quantitatively the original commutative cosmological model.
In particular, the NC parameter $\gamma$ acts as an additional repulsive force to the one already present in the model, 
due to the phantom fluid. That behavior happens for both $\gamma$ positive or negative. Therefore, the introduction of 
the present noncommutativity does not prevent the Universe ending in a {\it big rip} singularity. In fact, the {\it big rip}
singularity is reached, after the beginning of the expansion, first in the NC models than in the corresponding commutative 
ones. Since we are particularly interested in describing the present expansion of our Universe, we may mention that, due to 
the noncommutativity introduced here, we have an extra free parameter $\gamma$, not present in the corresponding
commutative models. One may use that extra freedom to better adjust the observational data. 

We also conclude that the NC scale factor behaves very much like the commutative one, when we vary most of the free
parameters and initial conditions of the NC model. When we increase the values of: $C$ (fluid energy), $a_0$ (initial 
scale factor value) and $\dot{a}_0$ (initial scale factor velocity), the NC scale factor goes quicker to the {\it big 
rip} singularity, like in the commutative case. It also goes quicker to the {\it big rip} singularity, when we diminish the
values of: $\alpha$ (negative parameter that defines the phantom fluid) and $k$ (parameter that gives the curvature of the
spatial sections), like in the commutative case. For $T_0$ (initial value of fluid variable $T$), which is not explicitly 
present in the commutative scale factor equation, the behavior of $a_{nc}$ is different for $\gamma$ positive or negative. 
For $\gamma > 0$, the greater the value of $T_0$, the more quickly the noncommutative scale factor reaches the {\it big rip}
singularity. On the other hand, for $\gamma < 0$, the greater the value of $T_0$, the more slowly the noncommutative scale
factor reaches the {\it big rip} singularity.

From our estimates for $\gamma$ and the time until the Universe reach the {\it big rip} singularity $t_{br}$ (Table \ref{comparison}), 
we can draw the following conclusions. The estimate values of $\gamma$ are very small. $\gamma$ increases 
as the time, the accelerated expansion started, approaches the initial moments of the Universe. That result is expected since
noncommutativity should had been more important at the beginning of the Universe. Due to the fact that the estimated values 
of $\gamma$ are very small, specially in comparison with the ones of $a_h$ (scale factor when the universe starts the present 
accelerated expansion), we observe that the time to reach the {\it big rip} ($t_{br}$) increases when the Universe starts to 
expand in an accelerated rate further back in time. 

{\bf Acknowledgements}. A. R. Vaz thanks CAPES for his scholarship.

\appendix

\section{Noncommutative Friedmann equation}

In the present appendix we write the NC Friedmann equation in terms of $\dot{a}_c$, $a_c$ and $T_c$. That equation 
is very important in the study of the solutions to the system Eqs. (\ref{13})-(\ref{14}). In order to do that, let
us write, initially, the NC superhamiltonian Eq. (\ref{6}), to first order in $\gamma$,
\begin{equation}
\label{a1}
\mathcal{H}_{nc} = - \frac{\left(P_{ac}^2 + \gamma P_{Tc}P_{a_{c}}\right)}{24} + \frac{P_{Tc} + \frac{\gamma P_{ac}}{2} }{a_{c}^{3\alpha - 1} + (\frac{3\alpha - 1}{2}) \gamma T_{c} a_{c}^{3\alpha - 2}} - 6k(a^2_{c}+2\gamma a_{c} T_{c}).
\end{equation}
Now, introducing $P_{Tc}$ Eq. (\ref{11}) and $P_{ac}$ Eq. (\ref{12}), in Eq. (\ref{a1}), we obtain, to first order in $\gamma$,
\begin{eqnarray}
\label{a2}
\mathcal{H}_{nc} & = & - \frac{\left( 144\dot{a}^2_{c} - 24\dot{a}_{c}\left( -\frac{\gamma C}{2} + 6\gamma \dot{T}_{c} \right) - 12\dot{a}_{c}\gamma C \right)}{24} \nonumber \\ 
&+&\frac{C - 12\gamma \dot{a}_{c} }{a_{c}^{3\alpha - 1} + (\frac{3\alpha - 1}{2}) \gamma T_{c} a_{c}^{3\alpha - 2}}
- 6k(a^2_{c}+\gamma a_{c} T_{c}).
\end{eqnarray}
Finally, introducing $\dot{T}_c$ Eq. (\ref{14}) in Eq. (\ref{a2}) and setting the resulting equation to zero, we obtain the Friedmann
equation in terms of $\dot{a}_c$, $a_c$ and $T_c$,
\begin{equation}
\label{a3}
6\gamma \dot{a}_{c} a_{c}^{1 - 3\alpha} - 6\dot{a}_{c}^2 + \frac{C - 12 \gamma \dot{a}_{c}}{a_{c}^{3\alpha - 1} + \frac{3\alpha - 1}{2}\gamma T_{c} a_{c}^{3\alpha - 2}} - 6k(a_{c}^2 +  \gamma a_{c} T_{c}) = 0.
\end{equation}

\end{document}